\documentclass{appolb}
\usepackage{epsfig}
\begin{document}
%\date{\today}
\pagestyle{plain}
%% uncomment the following line to get equations numbered by (sec.num)
%\eqsec
\newcount\eLiNe\eLiNe=\inputlineno\advance\eLiNe by -1
\title{$X(3872)$ as a $D\bar{D}^{*}$ MOLECULE BOUND BY \\
QUARK EXCHANGE FORCES%
%\thanks{Send any remarks to {\tt pena.ift.uni.wroc.pl}}%
}
\author{Carlos Pe\~na$^1$ and David Blaschke$^{1,2}$
\address{$^1$Institute for Theoretical Physics, University of Wroclaw,
%Max Born pl. 9, 50-204 Wroclaw,
Poland\\
$^2$Bogoliubov Laboratory for Theoretical Physics,
%Joint Institute for Nuclear Research, 141980 Dubna,
JINR Dubna, Russia}}
\maketitle

\begin{abstract}
The Bethe-Salpeter equation for the T-Matrix of $D^0\bar{D}^{*0}$
scattering is solved with a meson-meson potential that results from
$2^{\rm nd}$ order Born approximation of quark exchange processes.
This potential turns out to be complex and energy dependent due to the pole
contribution from the coupling to the intermediate $J/\psi$ - $\rho$
meson pair propagator.
As a consequence, a bound state with a mass close to $3.872$ GeV
occurs in the $J/\psi$ - $\rho$ continuum.
This result suggests that quark exchange forces may provide the solution
to the puzzling question for the origin of the interaction which leads to  a
binding of $D$ and $\bar{D}^{*}$ mesons in the $X(3872)$ state.
\end{abstract}

\section{Introduction}
The $X(3872)$ resonance was detected by Belle \cite{hep-ex/0309032}
by examining the invariant mass distribution of particles produced in the
decay of $B^{+}$ into  $K^{+}\;\pi^{+}\;\pi^{-}\;J/\psi$.
This sighting was later confirmed by  BaBar \cite{Aubert:2008gu}.
Even though the particle composition is still under discussion, this resonance
is likely to be a $D\bar{D}^{*}$ bound state with binding energy below
1~MeV \cite{Brambilla:2010cs}.
Some attempts have been presented to explain $X(3872)$ based on a T-matrix
approach taking into account the neutral and charged $D$ meson channels to
properly study the isospin violation in ${X}$ decay to $J/\psi+\pi^++\pi^-$ or
$J/\psi+\pi^++\pi^-+\pi^0$ \cite{Gamermann:2009uq,arXiv:1110.3694}.

In this contribution we present the derivation of a potential for the
$D\bar{D}^{*}$ interaction which is based on an extension of
the separable quark exchange interaction in the $D\bar{D}^{*}\to J/\psi+\rho$
channel \cite{Martins:1994hd} to the 2$^{\rm nd}$ order Born approximation
\cite{penaThesis:2011,xpena:2011}.
An elucidation of the nature of the $X(3872)$ state is important also from the
point of view of quark-gluon plasma search since it has been conjectured that
in a heavy-ion collision the $c\bar{c}$ state in {\it statu nacendi} may be
subject to important modifications due to its coupling to the $X(3872)$
resonance at the continuum threshold \cite{arXiv:1103.3155,arXiv:1106.2519}.
A generalization \cite{arXiv:1108.4180} of the Matsui model \cite{MIT-CTP-1756}
has been applied to the description of
the threshold-like structure in the very precise data from In-In
collisions at CERN-NA60 \cite{arXiv:1106.2519}.
Its relationship to the theory of the plasma Hamiltonian for heavy quarkonia
\cite{arXiv:0912.4479,arXiv:0807.2470}
is currently being worked out \cite{pena:2011}.
As a partial result of these studies we discuss here the role of quark exchange
processes as a possible binding mechanism leading to the $X(3872)$ as a
$D\bar{D}^{*}$ molecule.

\section{T-matrix Approach}
\begin{figure}[tbhp]
\begin{center}
\includegraphics[width=12cm,height=1.7cm]{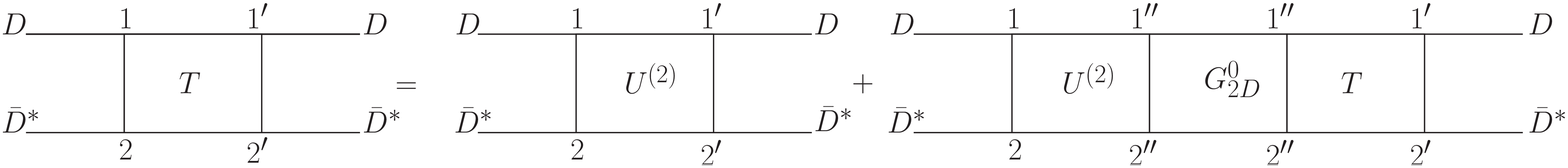}\vspace{-0.2cm}
\caption{Diagrammatic representation of Bethe-Salpeter equation for T-matrix of
$D\bar{D}^{*}$ scattering.
%Momentum coordinates connected by internal lines should be summed.
\label{Tmatrix}}
\end{center}
\end{figure}

The Bethe-Salpeter equation for a T-matrix description of $D\bar{D}^{*}$
scattering is written in the ladder approximation  as \cite{xpena:2011}
{\small
\begin{eqnarray}
\label{Tmatrix1}
T(a,a',z) &=& U^{(2)}(a,a')
+ \sum\limits_{a''}U^{(2)}(a,a'')\;G^0_{2D}(a'',z)\;T(a'',a',z)~,
 \end{eqnarray}}
which follows from the diagrammatic representation depicted in
Fig.~\ref{Tmatrix}.

For convenience we introduced the shorthand notation
$a=\textbf{\textsf{p}}_{1},\textbf{\textsf{p}}_{2}$,
$a'=\textbf{\textsf{p}}_{1^{'}},\textbf{\textsf{p}}_{2^{'}}$,
$a''=\textbf{\textsf{p}}_{1^{''}},\textbf{\textsf{p}}_{2^{''}}$,
$a'''=\textbf{\textsf{p}}_{1^{'''}},\textbf{\textsf{p}}_{2^{'''}}$.
Here $z$ refers to the energy of the scattering process described by
(\ref{Tmatrix1}).

\begin{figure}[tbhp]
\begin{center}
\includegraphics[width=12cm]{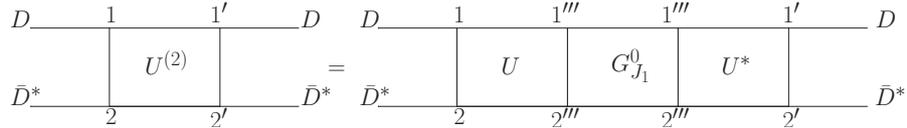}\vspace{-0.2cm}
\caption{Diagrammatic representation of the quark exchange interaction kernel
for $D\bar{D}^{*}$ scattering. The subscript $J_1$ stands for the channel
$J/\psi + \rho$.
%Momentum coordinates connected by internal lines should be summed.
\label{fig:U2}}
\end{center}
\end{figure}

For the interaction kernel $U^{(2)}$ we suggest here an extension of
first Born order diagrams for quark exchange processes in meson-meson
scattering \cite{MIT-CTP-2026,CERN-TH-6639-92}
to second order (see Fig.~\ref{fig:U2}), where the first order
process is given by a separable potential \cite{CERN-TH-6639-92,Martins:1994hd}
{\small
\begin{eqnarray}
\label{U}
\textbf{U}(a,a')&=&-\lambda\;L(a)\;R(a')~,~~
%\nonumber\\
\textbf{U}^{*}(a',a)=-\lambda\;R(a')\;L(a)~,
\end{eqnarray}}
where $L$ and $R$ are the meson form factors, dimensionless functions
interpolating between $1$ at zero relative momentum and $0$ at high momentum.
The amplitude  $\lambda$ of the interaction potential has the dimension of the
potential in momentum space, i.e. $[\lambda]=\textrm{GeV}^{-2}$.
The origin of such interaction can be explained by quark exchange forces
\cite{xpena:2011}.
With this ansatz the meson-meson potential in the $2^{\rm nd}$ Born
approximation (Fig.~\ref{fig:U2}) leads to a dynamic, separable potential
{\small
\begin{eqnarray}
\label{U2}
 \textbf{U}^{(2)}(a,a',z) &=& \displaystyle\sum\limits_{a'''}\textbf{U}(a,a''')\;G^0_{J_1}(a''',z)\;\textbf{U}^{*}(a''',a')\nonumber\\
 &=&\;L(a)\;L(a')\;\underbrace{\lambda^2\displaystyle\sum\limits_{a'''}R^2(a''')\;G^{0}_{J_1}(a''',z)}_{\textbf{\textrm{V}}(z)}\nonumber\\
  &=&\textbf{\textrm{V}}(z)\;L(a)\;L(a')~.
  \end{eqnarray}}
For a separable kernel, the Bethe-Salpeter equation for the T-matrix has
a solution in the form
{\small\begin{eqnarray}\label{Tmatrixelement}
T(a,a',z)=L(a)\;L(a')\;t(P,z)~,
\end{eqnarray}}
where $P$ is the total conserved momentum of the pair $D\bar{D}^{*}$.
By replacing (\ref{U}),  (\ref{U2})  and (\ref{Tmatrixelement}) into
(\ref{Tmatrix}) we find the solution
{\small
\begin{eqnarray}
\label{tmatrix}
t(P,z)=\frac{\textbf{\textrm{V}}(z)}{1-\underbrace{\textbf{\textrm{V}}(z)\;\displaystyle\sum\limits_{a''}L^{2}(a'')\;G^0_{2D}(a'',z)}_{B(P,z)}}~. \end{eqnarray}}
In the following we use the rest frame with vanishing total momentum ($P=0$)
and the relative momenta $p$, $p'$, where
$  \textbf{\textsf{p}}_{1^{''}}= p~$,
$  \textbf{\textsf{p}}_{2^{''}}= -p~$,
$  \textbf{\textsf{p}}_{1^{'''}}= p'~$,
$  \textbf{\textsf{p}}_{2^{'''}}=- p'~$.
%
%{\small\begin{eqnarray}
%  \textbf{\textsf{p}}_{\textbf{1}^{''}} &=& p~, \nonumber\\
%  \textbf{\textsf{p}}_{\textbf{2}^{''}} &=& -p~,\nonumber\\
%  \textbf{\textsf{p}}_{\textbf{1}^{'''}}&=& p'~, \nonumber\\
%  \textbf{\textsf{p}}_{\textbf{2}^{'''}} &=&- p'~.\nonumber\\
%\end{eqnarray}}
The function $G^0_{2D}(p,z)$ stands, depending on the considered channel,
for one of the nonrelativistic two-particle propagators $G^0_{D_1}(p,z)$ or 
$G^0_{J_1}(p,z)$, given by
{\small\begin{eqnarray}\label{propagators}
%G^0_{2D}(p,z)&=&G^0_{2D_1}(p,z)~,\nonumber\\
G^0_{D_1}(p,z)&=&\frac{1}{E_{D_1}-\frac{p^2}{2\mu_{D_1}}\pm i\varepsilon}~,~~
%\nonumber\\
%  G^0_{2D_2}(p,z)&=&
%\frac{1}{z-m_{D^{+}}-m_{\bar{D}^{*-}}-\frac{p^2}{2\mu_{D_2}}\pm i \varepsilon}
%~,\nonumber\\
G^0_{J_1}(p,z)=\frac{1}{E_{J_1}-\frac{p^2}{2\mu_{J_1}}\pm i \varepsilon}~.
%\nonumber\\
\end{eqnarray}}
The abbreviations $D_1=D^{0}+\bar{D}^{*0}$ and $J_1=J/\psi + \rho$ stand for 
the two-meson channels and $\mu_{D_1}$, $\mu_{J_1}$ for the corresponding 
reduced masses.
In the same framework the binding energies are defined by
{\small\begin{eqnarray}
  E_{D_1}&=&z-m_{D^{0}}-m_{\bar{D}^{*0}}~,~~
%\nonumber\\
  E_{J_1}=z-m_{J/\psi}-m_{\rho}~.
\end{eqnarray}}
We are interested in finding a resonance with a mass just below the threshold
of the $D_1$ continuum, in the region where
$m_{J/\psi}+m_{\rho}\leq z\leq m_{D^{0}}+m_{\bar{D}^{*0}}$.
The potential (\ref{U2}) turns out complex by considering
the pole contribution of the two-particle propagator $G^0_{J_1}(p,z)$.
This pole located at $p_{J_1}=\sqrt{2\mu_{J_1}E_{J_1}}$  provides the
meson-meson potential with sufficient strength to allow the formation of a
bound state. Thus integrating around the pole leads to
{\small
\begin{eqnarray}\label{v}
 \textbf{\textrm{V}}(z)&=&\lambda^2\displaystyle\sum\limits_{a'''}R^2(a''')\;G^{0}_{J_1}(a''',z)
%\nonumber\\&=&
=\frac{\lambda^2 \mu_{J_1}}{\pi^2}\;\displaystyle\lim_{\varepsilon\to 0}
\int^{\infty}_0dp\;\frac{p^2\;R^2(p)}{p^2_{J_1}-p^2 \pm i\varepsilon}
\nonumber\\
&=&\frac{\lambda^2 \mu_{J_1}}{\pi^2}\int^{\infty}_0 dp\;p^2\;R^2(p)
\Biggl[\frac{\wp}{p^2_{J_1}-{p^2}}
\mp i\pi\;\delta\left(p^2_{J_1}-{p^2}\right)\Biggl]
\nonumber\\
&=&\lambda^2\Biggl[\underbrace{\frac{\mu_{J_1}}{\pi^2}
\displaystyle\lim_{\varepsilon\to 0}\int^{\infty}_0dp\;
\frac{p^2\;R^2(p)(p^2_{J_1}-{p^2})}{(p^2_{J_1}-{p^2})^2+\varepsilon^2}}_{
A_{J_1}(z)}
+ i\;\underbrace{\left(\mp\frac{\mu_{J_1}\;p_{J_1}\;R^2(p_{J_1})}{2\pi}\right)
\Theta(E_{J_1})}_{B_{J_1}(z)}\Biggl]~.
\nonumber\\
   \end{eqnarray}}
Similarly,  the integration around {\small$p_{D_1}=\sqrt{2\mu_{D_1} E_{D_1}}$}
gives
{\small\begin{eqnarray}\label{B}
B(0,z)&=&\textbf{\textrm{V}}(z)\frac{\mu_{D_1}}{\pi^2}\;\int^{\infty}_0dp\;
\frac{p^2\;L^{2}(p)}{p^2_{D_1}-{p^2}\pm i\varepsilon}
\nonumber\\
&=&\textbf{\textrm{V}}(z)\Biggl[\underbrace{\frac{\mu_{D_1}}{\pi^2}
\displaystyle\lim_{\varepsilon\to 0}\int^{\infty}_0dp\;
\frac{p^2\;L^2(p)(p^2_{D_1}-{p^2})}{(p^2_{D_1}-{p^2})^2+\varepsilon^2}}_{
A_{D_1}(z)}
+ i\;\underbrace{\left(\mp\frac{\mu_{D_1}\;p_{D_1}\;L^2(p_{D_1})}{2\pi}\right)
\Theta(E_{D_1})}_{B_{D_1}(z)}\Biggl]~.
\nonumber\\
\end{eqnarray}}
Notice that the step function $\Theta(E_{D_1})$ cancels the term
$B_{D_1}(z)$ for $E_{D_1}\leq 0$.

\section{Scattering phase shift}
\begin{figure}[tbhp]
\begin{center}
\includegraphics[width=6cm]{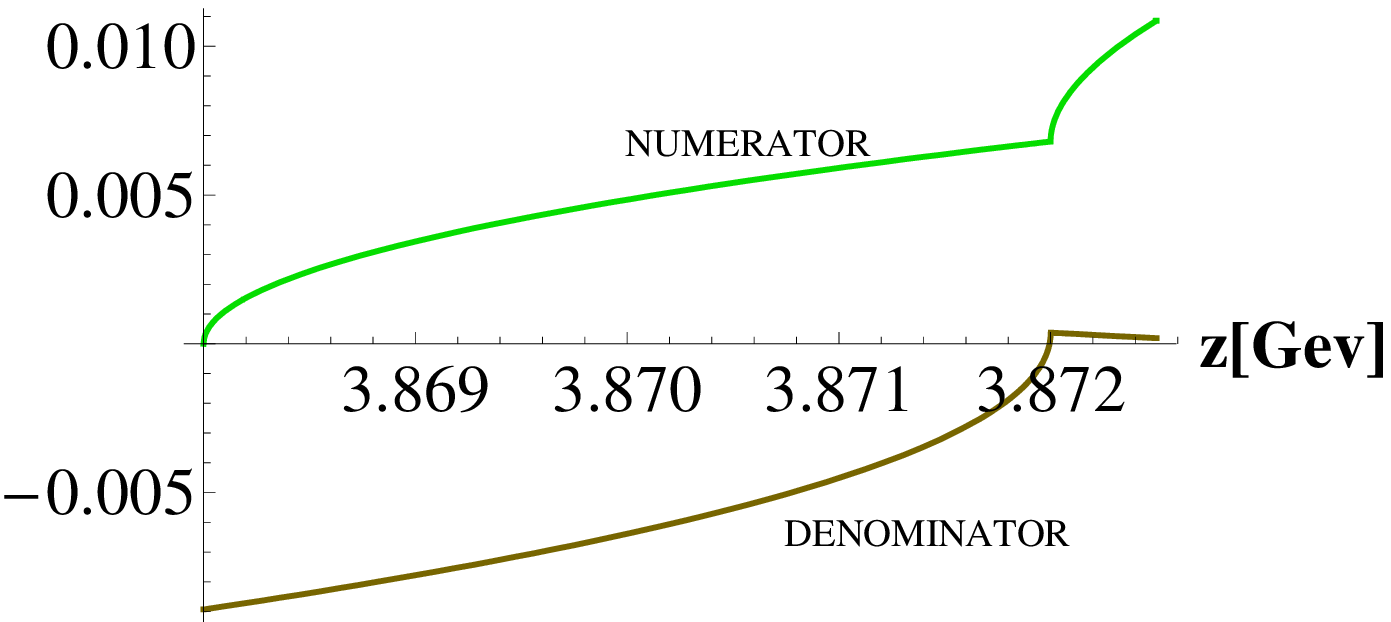}
\includegraphics[width=6cm]{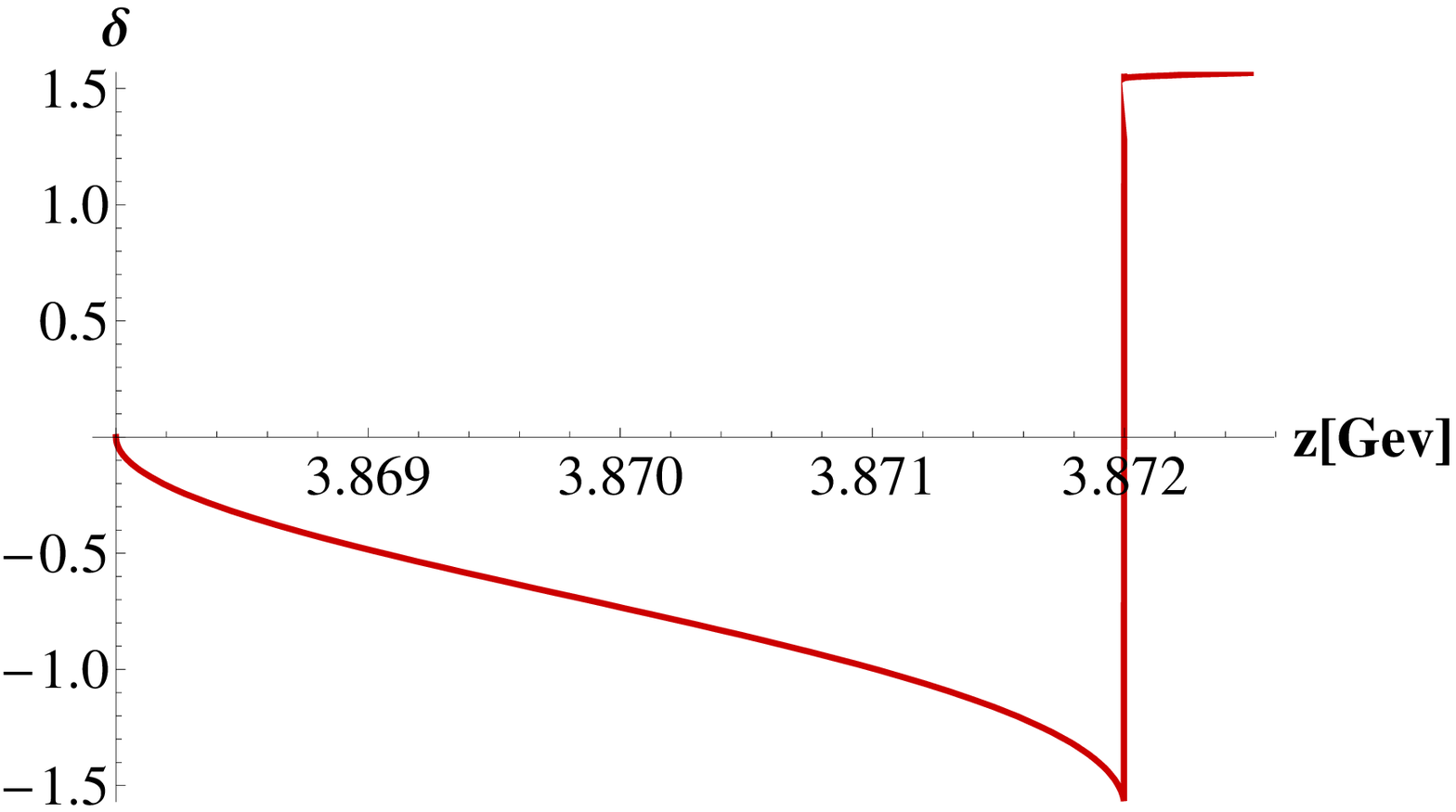}
\caption{Plots are done by choosing
$\lambda=20.3~\textrm{GeV}^{-2}$, $\gamma=0.8$ GeV and sign (+) for imaginary
parts.
{Left Panel}:
Numerator and denominator of expression (\ref{phaseshift}).
Notice the behavior of the denominator just around the threshold of $D$ mesons
which vanishes exactly at $z=3.872$ GeV.
{Right Panel}:
The sharp jump of the scattering phase shift by $\pi$ indicates that a
quasi-bound state (resonance) with a mass of $z=3.872$ GeV appeared.
A similar example of such characteristic behavior can be observed in the
deuteron channel in nuclear matter: just above the critical temperature for
Bose condensation of Cooper pairs a quasi-bound state appears in the
continuum, see Fig.~6 of Ref.~\cite{Schmidt199057}.
\label{fig:ND1}}
\end{center}
\end{figure}

The scattering phase shift is determined as
\begin{eqnarray}\label{phaseshift}
\tan(\delta) &=& \frac{Im[t(0,z)]}{Re[t(0,z)]}
%\nonumber\\&=&
=\frac{B_{J_1}(z)+\left(A^2_{J_1}(z)+B^2_{J_1}(z)\right) B_{D_1}(z) \lambda^2}
{A_{J_1}(z)-\left(A^2_{J_1}(z)+ B^2_{J_1}(z)\right)A_{D_1}(z) \lambda^2}~.
\end{eqnarray}
The expression (\ref{phaseshift}) is very useful to decide whether or not a
bound state is located in the region
$m_{J/\psi}+m_{\rho}\leq z\leq m_{D^{0}}+m_{\bar{D}^{*0}}$.
This occurs when $B_{J_1}(z)\neq 0$.
Therefore, the analytical continuation of the meson-meson interaction is
crucial for explaining resonance formation.
An instructive calculation is performed with a Lorentzian form factor
(Yamaguchi potential with width parameter $\gamma$), using the integral
{\small
$\int^\infty_0 dx\frac{x^2}{(x^2+1)^2(x^2-\tilde{E})}=
\frac{\pi}{4}\frac{1}{(1+\sqrt{-\tilde{E}})^2}$}
with $\tilde{E}=\frac{2\mu E}{\gamma^2}$.
Notice that for $E>0$ the equality splits into real and imaginary
parts, otherwise it is always real \cite{Ropke:2009}.
The result is
{\small\begin{eqnarray}
B_{J_1}(\tilde{E}_J)&=&\mp\frac{\mu_J\gamma}{2\pi}
\frac{\sqrt{\tilde{E}_J}}{(1+\tilde{E}_J)^2}\theta(\tilde{E}_J)~,~~~
%\nonumber\\
B_{D_1}(\tilde{E}_{D_1})=\mp\frac{\mu_{D_1}\gamma}{2\pi}
\frac{\sqrt{\tilde{E}_{D_1}}}{(1+\tilde{E}_{D_1})^2}\theta(\tilde{E}_{D_1})~,
\nonumber\\
A_{J_1}(\tilde{E}_J)&=&-\frac{\mu_{J_1}\gamma}{4\pi} \left\{ \begin{array}{ll}
\frac{1-\tilde{E}_J}{(1+\tilde{E}_J)^2}& \textrm{$\tilde{E}_J> 0~,$}\\
\frac{1}{(1+\sqrt{-\tilde{E}_J})^2}& \textrm{$\tilde{E}_J\leq 0~,$}\\
\end{array} \right.
\nonumber\\
A_{D_1}(z)&=&- \frac{\mu_{D_1}\gamma}{4\pi} \left\{ \begin{array}{ll}
\frac{1-\tilde{E}_{D_1}}{(1+\tilde{E}_{D_1})^2}& \textrm{$\tilde{E}_{D_1}> 0~,$
}\\
\frac{1}{(1+\sqrt{-\tilde{E}_{D_1}})^2}& \textrm{$\tilde{E}_{D_1}\leq 0~.$}\\
\end{array} \right.
\end{eqnarray}}
For a particular choice of $\lambda$ and $\gamma$ it is possible that the
denominator of the expression (\ref{phaseshift}) vanishes at threshold
$z=m_{D^{0}}+m_{\bar{D}^{*0}}=3.872$ GeV. The interesting features of these results are shown in Fig.~\ref{fig:ND1}
where the scattering phase shift reflects the production of a quasi-bound state
by mean interaction of two D-mesons, $D^{0}$ and $\bar{D}^{*0}$, passing through the $J/\psi+\rho$
channel.
\section{Conclusions}
In this work we have considered the solution of the Bethe-Salpeter equation
for the T-matrix of $D^0\bar{D}^{*0}$ scattering which via quark exchange
couples to the virtual propagation of a $J/\psi-\rho$ pair.
The kernel of the Bethe-Salpeter equation is a complex meson-meson potential
obtained by a $2^{\rm nd}$ order Born approximation to quark exchange.
We have shown that this analytical continuation is crucial for the formation
of a bound state in the region
$m_{J/\psi}+m_{\rho}\leq z\leq m_{D^{0}}+m_{\bar{D}^{*0}}$.
The dynamical nature of this complex potential provides a sufficient
enhancement of the strength at the threshold which leads to the $X(3872)$
bound state, rather independent of the detailed dynamics of the model.
In a next step, we consider the coupled channel problem, including charged $D$
meson states and the $J/\psi-\omega$ channel \cite{xpena:2011}.
%\\[1cm]
\subsection*{Acknowledgments}
C.P. acknowledges a grant of the Polish Government for supporting part of his 
work. D.B. was supported in part by the Polish Ministry for Science and Higher 
Education under grant No. NN 202 2318 37 and by the Russian
Foundation for Basic Research under grant No. 11-02-01538-a.

%%%%%%%%%%%%%%%%%%%%%%%%%%%
% References              %
%%%%%%%%%%%%%%%%%%%%%%%%%%%

\end{document}